\documentclass[10pt,conference,a4paper]{IEEEtran}
\usepackage{epsfig}
\usepackage{cite}
\usepackage{graphicx}
\usepackage[cmex10]{amsmath}
\usepackage{url}

\usepackage{multirow}
\usepackage{epstopdf}

\begin{document}

\title{A Convolutional Neural Network Approach for Half-Pel Interpolation in Video Coding}

\author{
\IEEEauthorblockN{Ning Yan, Dong Liu*, Houqiang Li, and Feng Wu}
\IEEEauthorblockA{CAS Key Laboratory of Technology in Geo-Spatial Information Processing and Application System,\\
University of Science and Technology of China, Hefei 230027, China\\
\texttt{nyan@mail.ustc.edu.cn, \{dongeliu,lihq,fengwu\}@ustc.edu.cn}}
}

\maketitle
\begin{figure}[b]
\fontsize{8}{9}\selectfont
*Corresponding author. This work was supported by the National Key Research and Development Plan under Grant 2016YFC0801001, by the Natural Science Foundation of China (NSFC) under Grants 61390512, 61331017, and 61632001, and by the Fundamental Research Funds for the Central Universities under Grant WK3490000001.
\end{figure}

\begin{abstract}
Motion compensation is a fundamental technology in video coding to remove the temporal redundancy between video frames.
To further improve the coding efficiency, sub-pel motion compensation has been utilized, which requires interpolation of fractional samples.
The video coding standards usually adopt fixed interpolation filters that are derived from the signal processing theory.
However, as video signal is not stationary, the fixed interpolation filters may turn out less efficient.
Inspired by the great success of convolutional neural network (CNN) in computer vision, we propose to design a CNN-based interpolation filter (CNNIF) for video coding.
Different from previous studies, one difficulty for training CNNIF is the lack of ground-truth since the fractional samples are actually not available.
Our solution for this problem is to derive the ``ground-truth'' of fractional samples by smoothing high-resolution images, which is verified to be effective by the conducted experiments.
Compared to the fixed half-pel interpolation filter for luma in High Efficiency Video Coding (HEVC), our proposed CNNIF achieves up to 3.2\% and on average 0.9\% BD-rate reduction under low-delay P configuration.
\end{abstract}

\IEEEpeerreviewmaketitle

\section{Introduction}
Motion compensation is a fundamental technology in video coding to remove the temporal redundancy between video frames.
The existing video coding standards, including High Efficiency Video Coding (HEVC), mostly adopt block-based motion compensation for inter prediction, which assumes the block to be coded can be retrieved from the previously coded frames, and the corresponding location of the retrieved block is indicated by a motion vector (MV).
Due to the inherent spatial sampling of digital video, MV is probably not integer, and thus the corresponding block may need to be generated instead of simply retrieved.
Typically, generating non-integer samples is performed by kinds of interpolation, which can be viewed as fitting a continuous curve through a set of discrete samples, and picking out the new values at specific positions on the curve.

The problem of fractional sample interpolation for video coding has been widely studied \cite{MOMS}.
A theoretical analysis is conducted in \cite{Girod} about the influence of fractional-pel accuracy on the efficiency of motion compensated prediction, using a Gaussian power spectral density model.
Practically, the video coding standards mostly adopt fixed interpolation filters. For example, MPEG-4 AVC/H.264 uses the 6-tap filter to perform half-pel interpolation and the simple average filter to perform quarter-pel interpolation for luma component \cite{wiegand2003overview}. In HEVC, the DCT-based interpolation filter (DCTIF) is adopted \cite{sullivan2012overview}. HEVC uses a ``7q+8h'' DCTIF for luma component, that is a 7-tap DCTIF used for quarter-pel samples, and a 8-tap DCTIF for half-pel samples. Lv \emph{et al.} \cite{DCTIF} give the derivation process of DCTIF in detail, and compare the frequency responses between the interpolation filters in HEVC and H.264.

The fixed interpolation filters have been designed according to the signal processing theory, with the premise that the video signal is an ideal low-pass one. However, the video signal is indeed not low-pass and not stationary. Further study has been conducted to design different interpolation filters. For example, a motion compensated hybrid video coding scheme using an adaptive filter is presented in \cite{wedi2001adaptive}, where the filter coefficients are estimated during motion compensation for each frame. Wittmann \emph{et al.} \cite{wittmann2008separable} present a separable adaptive interpolation filter to reduce the computational cost while maintaining the coding efficiency of non-separable adaptive filter. Nonetheless, the previous works adopted hand-crafted filters, leaving a space for further improving the accuracy of motion compensation.

Recently, deep learning has achieved great success in computer vision. Convolutional neural network (CNN) based models led to a series of breakthroughs in high-level computer vision tasks, such as image classification \cite{AlexNet} and object detection \cite{GoogleNet}. Later on, CNN is also utilized in some low-level computer vision tasks. For example, Dong \emph{et al.} \cite{dong2014learning} propose a CNN approach for image super-resolution, termed SRCNN, which learns an end-to-end mapping between low- and high-resolution images. SRCNN has achieved significant boost of performance in both subjective and objective quality, compared to the previous methods without CNN. SRCNN is then extended to cope with the problem of artifact reduction \cite{ARCNN}. More recently, Dai \emph{et al.} \cite{VRCNN} propose to learn a CNN for post-processing in video coding, and demonstrate on average 4.6\% bit-rate reduction than HEVC baseline. All these works seem to open up a new direction that adopts CNN into video coding to further improve the coding efficiency.

In this paper, we present a CNN approach for fractional sample interpolation in video coding. We expect the CNN to automate the discovery of interpolation filters rather than to design them manually. However, a key difficulty here is how to generate training data for CNN. In previous studies concerning CNN, the ground-truth labels are provided comprehensively in the training data. However, for fractional sample interpolation, we have no ground-truth label because the fractional samples actually do not exist. Our solution for this problem is to derive the labels by smoothing high-resolution images, which is verified to be effective in experiments. We then reuse the network architecture of SRCNN, considering the similarity between super-resolution and fractional interpolation, but train the network with our derived training data. After training, the CNN-based interpolation filter (CNNIF) is integrated into HEVC for testing its performance in video coding. Currently, the CNNIF is applied only to half-pel samples of luma component. Experimental results show that the proposed CNNIF leads to up to 3.2\% and on average 0.9\% bits saving under low-delay P configuration.

The remainder of this paper is organized as follows. Section II provides the details of the proposed CNNIF. Section III gives the experimental results, and Section IV concludes the paper.

\section{Proposed Method}
\subsection{Overview of the Proposed Method}

\begin{figure}
  \centering
  \includegraphics[width=65mm]{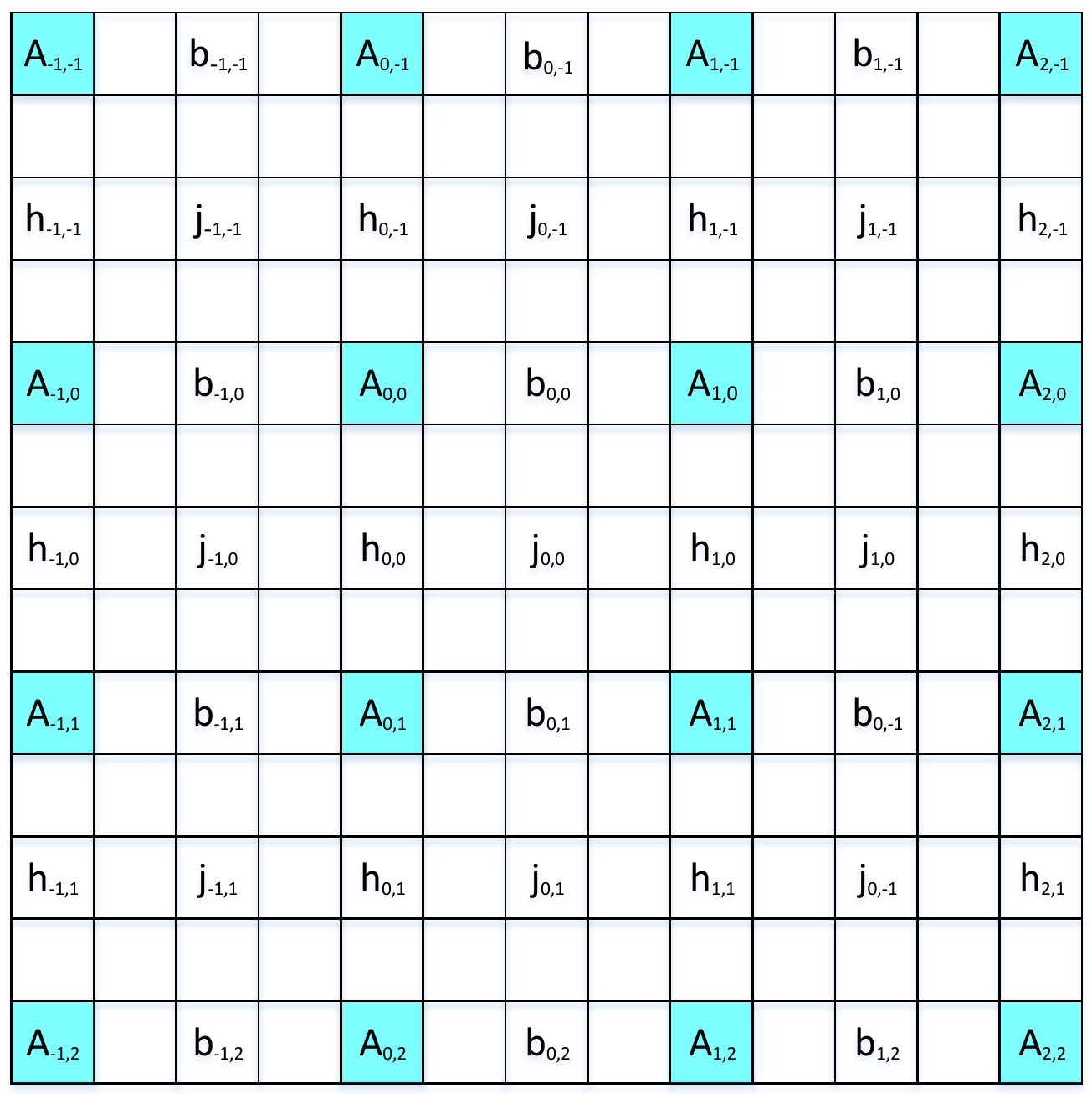}
  \caption{Integer and fractional sample positions during luma interpolation. A, b, h, and j stand for integer, horizontal half-pel, vertical half-pel, and diagonal half-pel positions, respectively.}
  \label{fig:pos}
\end{figure}

In this work, we propose a CNN approach for half-pel interpolation of luma component. Fig. \ref{fig:pos} illustrates the integer and half-pel positions during interpolation. The positions labeled with A$_{i,j}$ stand for the available luma samples at integer locations, whereas the other positions labeled with b$_{i,j}$, h$_{i,j}$ and j$_{i,j}$ represent samples at half-pel locations, which need to be interpolated from integer-location samples. In HEVC, these three positions are derived using uniform 8-tap DCTIF. Taking into account the positional relation between the integer and the half-pel samples, we propose to train three CNN models to perform the interpolation of horizontal half-pel, vertical half-pel, and diagonal half-pel samples, respectively. That is, we train CNNIF\_H, CNNIF\_V and CNNIF\_D, for the interpolation of b$_{i,j}$, h$_{i,j}$, and j$_{i,j}$, respectively.

It is very important to have abundant training data in store for supervised deep learning. However, in this fractional interpolation task, we cannot obtain the real ground-truth for training, i.e. the fractional samples do not exist actually. Without the ground-truth, it is not possible to carry out the training task, let alone expecting performance improvement in video coding. To overcome this contradiction, in this work, we design a method for generating training data, which will be introduced in Section II-C.

\subsection{Network Architecture}
\begin{figure}
  \centering
  \includegraphics[width=90mm]{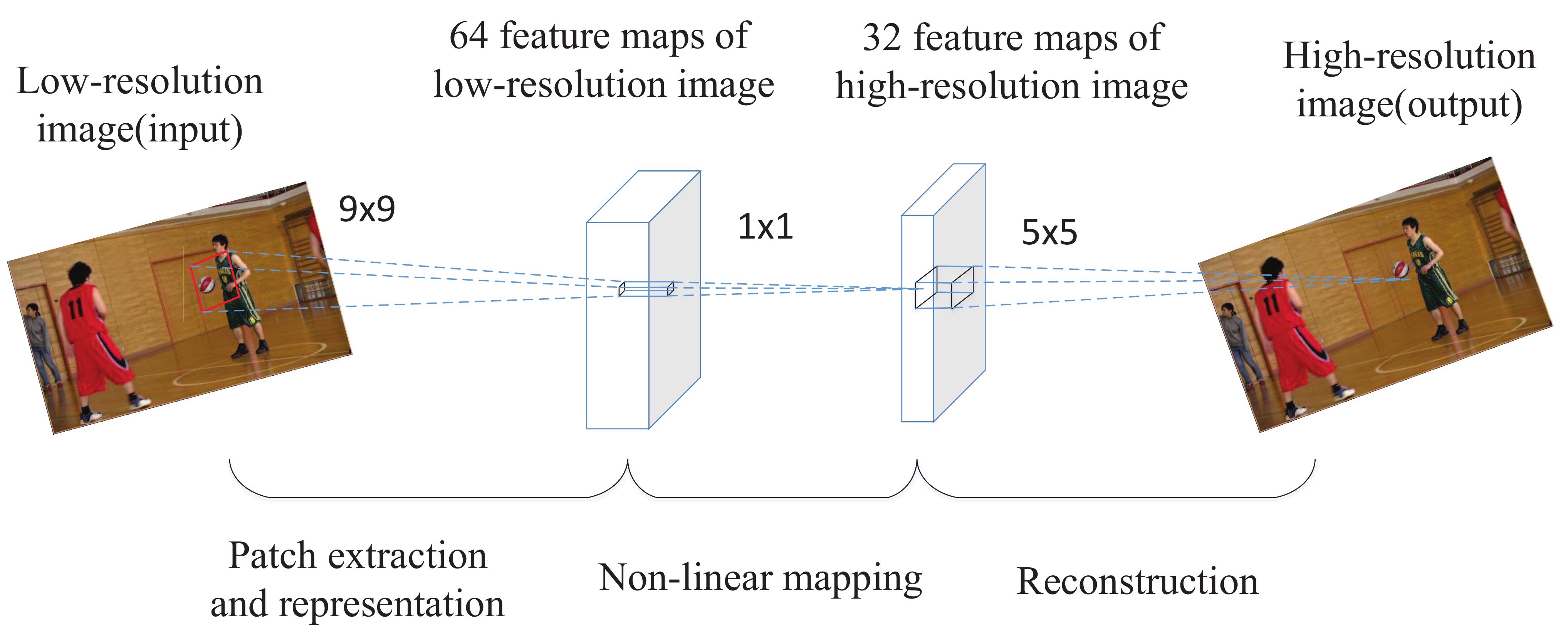}
  \caption{The architecture of super-resolution convolutional neural network (SRCNN) \cite{dong2014learning}, also used in this work.}
  \label{fig:arch}
\end{figure}

In this work, we reuse the existing Convolutional Neural Network for Super-Resolution (SRCNN) in \cite{dong2014learning} to carry out the half-pel interpolation task. Fig. \ref{fig:arch} depicts the architecture of SRCNN.

SRCNN consists of three convolutional layers. The output of $i$-th layer ($i=1,2$) is the result of a linear transform of the previous layer followed by a Rectified Linear Unit (ReLU) \cite{ReLU}, and this process can be expressed as:
$$
F_{i} = \max(0,W_{i} * F_{i-1}+B_{i})   \eqno{(1)}
$$
where $W_{i}$ and $B_{i}$ are the convolutional filter kernel and bias of the $i$-th layer, respectively, and `*' means convolutional operation. For SRCNN, the three layers are claimed to perform three steps, respectively \cite{dong2014learning}:

\begin{itemize}
\item The first layer is used for patch extraction and representation, extracting the features from low-resolution image. Here, $W_{1}$ is of size $9\times9\times 64$ and $B_{1}$ is a 64-dimensional vector.
\item The second layer can be seen as non-linear mapping, which converts the features of low-resolution image to those of high-resolution. Here, $W_{2}$ is of size $1\times 1\times 32$ and $B_{2}$ is a 32-dimensional vector.
\item The third layer, where $W_{3}$ is of size $5\times5\times1$, is used to recover the high-resolution image from the high-resolution features.
\end{itemize}

In the super-resolution task, a low-resolution image is firstly up-scaled to the desired size using bicubic interpolation, the interpolated image is denoted as $Y$. The goal is to recover an image $F(Y)$ to be as similar as possible to the ground-truth image $X$. While in our fractional interpolation task, the input of the network is the image consisting of integer-position pixels, denoted by $Y_{int}$, and the output is the interpolated image of fractional positions, $F_{h}$, which has the same size with the input image:
$$
F_{h} = W_{3}*F_{2} + B_{3}   \eqno{(2)}
$$

Please note that fractional sample interpolation is related to but different from super-resolution: the former tries to generate only fractional samples, while the latter is to generate a complete high-resolution image. If we simply reuse super-resolution for fractional sample interpolation, the integer-position samples cannot be guaranteed to be identical. Our experimental results also show that the simple reusing does not work well.
\subsection{Derivation of Training Data}
The derivation of training data is performed in two steps:

\begin{itemize}
\item Blurring a training image with a low-pass filter.
\item Extracting input and label for CNNIF\_H, CNNIF\_V, and CNNIF\_D, respectively.
\end{itemize}

\subsubsection{Image Blurring}

Like image super-resolution, fractional interpolation is also an ill-posed problem. One of the most difficult issues in training the CNN for fractional interpolation is the absence of ground-truth, since the fractional pixels are not available. It is infeasible for training the CNN if we do not have the ground-truth.

The generation of digital images is a process of sampling from the analog signal, which is essentially a process of low-pass filtering followed by decimation. Therefore, analogous to the derivation of digital images, we propose to firstly blur the training images with a low-pass filter to simulate the process of sampling. And intuitively, this operation can increase the correlation between neighboring pixels, so the relation between two neighboring pixels is more like that between integer and half-pel samples.

\subsubsection{Data Extracting}
\begin{figure}
  \centering
  \includegraphics[width=90mm]{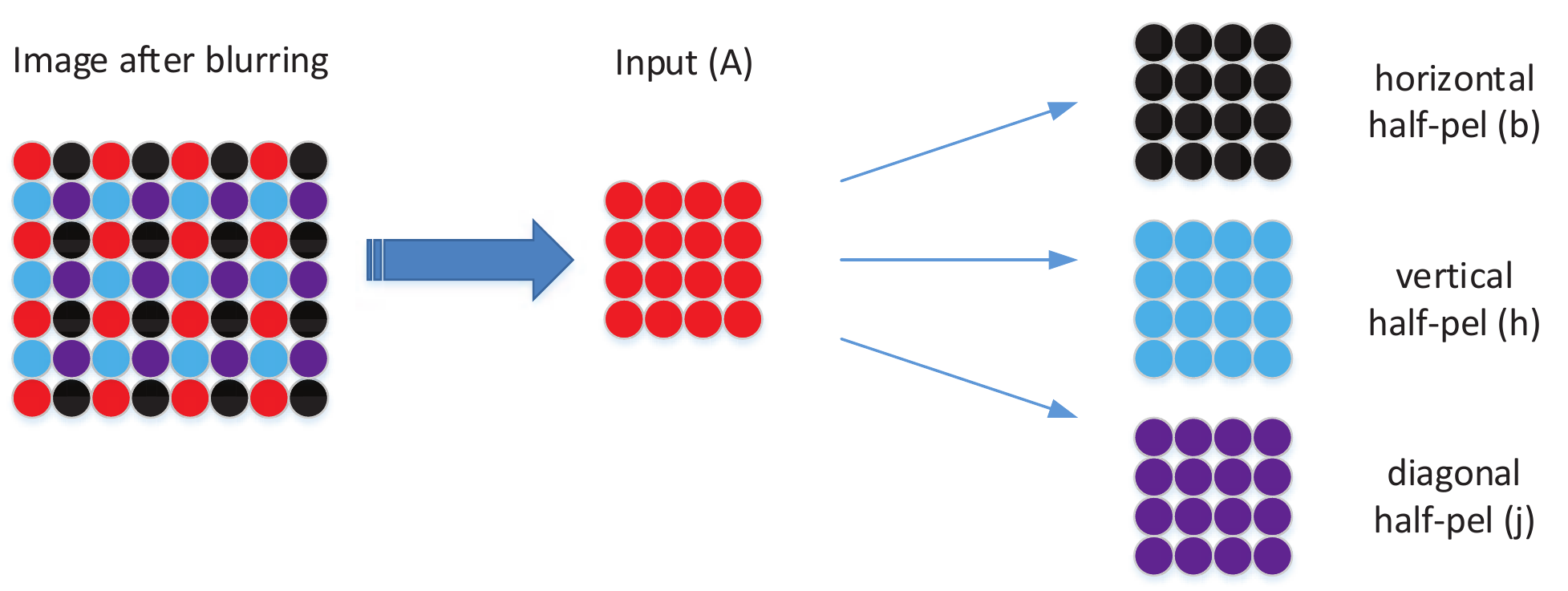}
  \caption{The process of generating labeled training data. A high-resolution image is blurred and then odd and even positions are regarded as integer and half-pel positions, respectively. A, b, h, and j correspond to those in Fig. \ref{fig:pos}.}
  \label{fig:gt}
\end{figure}

After blurring the training images, the correlation between the neighboring pixels of the images is more like that between integer and half pixels. Fig. \ref{fig:gt} shows the process of extracting input and labels for training. The pixels of phase zero (red points in the figure) are extracted as the input image of CNN and regarded as integer pixels. The pixels of phase one (black points) are used as the horizontal half pixels. Similarly, the pixels of phase two and phase three (green and purple points) are used as vertical and diagonal half pixels, respectively.

It is noticeable that the fractional interpolation is performed upon the reconstruction of the previously coded frames. For lossy video coding, especially with high quantization parameter (QP), significant reconstruction error will be introduced, and thus the interpolation accuracy will deteriorate. To correct this, the pixels of phase zero (red points in Fig. \ref{fig:gt}) are firstly coded and reconstructed by HEVC intra coding, and the reconstruction results are used as the inputs of CNNIF. In other words, we hope the CNNIF to generate the half-pel pixels from the compressed integer pixels.

Up to now, we have generated the training data of the three CNN models for half-pel interpolation. The pairs shown in Fig. \ref{fig:gt} (Input, horizontal half-pel), (Input, vertical half-pel) and (Input, diagonal half-pel) are used as the training data of CNNIF\_H, CNNIF\_V, and CNNIF\_D, respectively.

\subsection{Training Method}
In this task, the three CNN models are trained in the same way, and the loss function is optimized using stochastic gradient descent with back-propagation. The training of CNN is actually a process of adjusting the parameters set $\varTheta$, i.e. $(W_{1},W_{2},W_{3},B_{1},B_{2},B_{3})$ for SRCNN, to minimize the loss function over the training set. Let $F$ denotes the output of the CNNIF, and the labels are denoted as $\{Y^i\}_{i=1}^N$. Here, we use the Euclidean distance as the loss function:
$$
L = \frac{1}{N} \sum_{i=1}^{N}\|F^i-Y^i\|^2 \eqno{(3)}
$$
where $N$ is the total number of training data items.

\section{Experiments}

\subsection{Training Stage}
We use the deep learning framework Caffe \cite{Caffe} to train the CNNIFs on an NVIDIA Tesla K40C graphical processing unit (GPU). The training set we use is the same as that in \cite{ARCNN}, which is a collection of 400 natural images. All the images are processed using the method depicted in Fig. \ref{fig:gt}, and after that we have 400 sub-images as input and 1200 sub-images for the labels (400 sub-images for each CNNIF). In this implementation, each sub-image as input is compressed by HEVC intra coding at four different QPs: 22, 27, 32, and 37. For each QP and each half-pel position, a separate network is trained. Therefore, we finally train 12 CNNIFs. During the process of compression, a CNNIF will be selected according to the slice QP and the corresponding half-pel position. The nearest QP among 22, 27, 32, and 37 to the current slice QP will be considered.
\subsection{Comparison with HEVC Baseline}
The proposed method is implemented based on HEVC reference software HM 16.7. Currently, only the process of half-pel interpolation of the luma component is replaced by CNNIFs. The low delay P (LDP) configuration is tested in the experiment under the HEVC common test conditions. BD-rate is used to measure the rate-distortion (RD) performance. The experimental results are summarized in Table I. As can be observed, the proposed method achieves on average 0.9\% BD-rate reduction. For the test sequence \texttt{BQTerrace}, the BD-rate reduction can be as high as 3.2\%, 1.6\%, 1.6\% for Y, U, V components, respectively. Since the fractional interpolation of chroma components is still DCTIF, the performance of chroma components is not prominent. In the future work, we will train CNN models for the chroma components.

\begin{table}
\center
\caption{BD-Rate Results of Our CNNIF Compared to HEVC Baseline}
\begin{tabular}{|l|l|c|c|c|}
 \hline
  \multirow{2}{*}{Class}& \multirow{2}{*}{Sequence}&\multicolumn{3}{c|}{BD-rate}\\
 \cline{3-5}
 &&Y (\%) & U (\%) &V (\%)\\
 \hline
 \multirow{5}{*}{Class B} & Kimono & -1.1&0.1 &0.2\\
  \cline{2-5}
&ParkScene &-0.4 &-0.3 &-0.3\\
  \cline{2-5}
 &Cactus &-0.8&	0.0&	0.3\\
 \cline{2-5}
&BasketballDrive & -1.3	&-0.2	&-0.1\\
 \cline{2-5}
&BQTerrace & -3.2&	-1.6&	-1.6\\
 \hline
\multirow{4}{*}{Class C}& BasketballDrill &-1.2&	-0.6&	0.2\\
  \cline{2-5}
 &BQMall & -0.9&	0.2&	0.7\\
  \cline{2-5}
&PartyScene &0.2	&0.5	&0.3\\
\cline{2-5}
&RaceHorses &-1.5&	-0.5	&-0.1\\
 \hline
 \multirow{4}{*}{Class D} & BasketballPass & -1.3&	-0.4&	0.3\\
 \cline{2-5}
&BQSquare &  1.2&	2.9&	3.1\\
 \cline{2-5}
 &BlowingBubbles & -0.3&	0.4&	0.8\\
 \cline{2-5}
&RaceHorses & -0.8&	-0.9&	0.0\\
 \hline
\multirow{3}{*}{Class E}& FourPeople & -1.3&	-0.4&	0.1\\
 \cline{2-5}
 &Johnny & -1.2&	-0.4&	-0.7\\
 \cline{2-5}
&KristenAndSara & -1.0&	0.3&	0.2\\
\hline
\multirow{4}{*}{Class F}& BasketballDrillText & -1.4&	-0.2&	0.1\\
 \cline{2-5}
 &ChinaSpeed & -0.6&	-0.5&	-0.3\\
 \cline{2-5}
&SlideEditing & 0.0&	0.3&	0.4\\
 \cline{2-5}
 &SlideShow & -0.7&	-0.1&	-0.2\\
 \cline{2-5}
\hline
 \multirow{5}{*}{Class Summary}
 &Class B&-1.4&	-0.4&	-0.3\\
  \cline{2-5}
 &Class C&-0.9&	-0.1&	0.3\\
  \cline{2-5}
 &Class D&-0.3&	0.5&	1.0\\
  \cline{2-5}
 &Class E&-1.2&	-0.2&	-0.1\\
   \cline{2-5}
 &Class F&-0.7&	-0.1&	0.0\\
 \hline
\textbf{Overall} &\textbf{All}& \textbf{-0.9} & \textbf{-0.1}&\textbf{0.2}\\
 \hline

\end{tabular}
\end{table}

\subsection{Comparison with Super-Resolution}
We also compare our proposed method with super-resolution method. As an anchor, the reconstructed frame is up-scaled to a larger one by the pre-trained SRCNN model \cite{SRCNNModel} with a factor of 2. From the enlarged frame, the phase one, phase two and phase three pixels are used as the interpolated half-pel samples. We integrate this anchor method into HM 16.7. Table II shows the results of the HEVC test sequences Class C and Class D of the anchor method compared to HEVC baseline. It can be observed that all the sequences suffer from significant loss. For the test sequence \texttt{BQSquare}, the loss can be as high as 8.2\% for luma component. Therefore, despite the similarity between fractional interpolation and image super-resolution, they are indeed not the same task, and it is not appealing to directly apply super-resolution method to fractional interpolation, as validated by the experimental results.

\begin{table}
\center
\caption{BD-Rate Results of SRCNN \cite{SRCNNModel} Compared to HEVC Baseline}
\begin{tabular}{|l|l|c|c|c|}
 \hline
  \multirow{2}{*}{Class}& \multirow{2}{*}{Sequence}&\multicolumn{3}{c|}{BD-rate}\\
 \cline{3-5}
 &&Y (\%) & U (\%) &V (\%)\\
 \hline
\multirow{4}{*}{Class C}& BasketballDrill &0.8&	1.2&	2.1\\
  \cline{2-5}
 &BQMall & 2.8&	2.7&	3.0\\
  \cline{2-5}
&PartyScene &3.6	&3.4	&3.7\\
\cline{2-5}
&RaceHorses &2.4&	2.1	&2.0\\
 \hline
 \multirow{4}{*}{Class D} & BasketballPass & 1.7&	1.3&	2.0\\
 \cline{2-5}
&BQSquare &  8.2&	7.9&	6.8\\
 \cline{2-5}
 &BlowingBubbles & 3.2&	3.5&	4.2\\
 \cline{2-5}
&RaceHorses & 3.6&	2.0&	2.3\\
 \hline
\end{tabular}
\end{table}

\enlargethispage{-18.3mm}

\section{Conclusion}
This paper presents a convolutional neural network based fractional interpolation for inter prediction in HEVC. We use the existing SRCNN structure, but retrain interpolation CNN models for the three half-pel positions of luma component. A blurring followed by extracting method is proposed to generate training data, especially the missing labels. Experimental results show that the proposed CNNIF can achieve on avearge 0.9\% bits saving. Our further work will focus on two aspects. First, to design a more efficient network architecture that is more suitable for the interpolation task. Second, we will investigate how to generate better labels for training, especially for the quarter-pel interpolation.

\bibliographystyle{IEEEtran}

\bibliography{IEEEabrv,IEEEexample}

\end{document}